\documentstyle[prl,twocolumn,aps,floats,epsfig]{revtex}

\begin{document}

\draft
\twocolumn[\hsize\textwidth\columnwidth\hsize\csname 
@twocolumnfalse\endcsname

\title{Quasi-one-dimensional topological-excitation 
liquid in Bi$_{2}$Sr$_{2}$CaCu$_{2}$O$_{8+x}$ from tunneling 
spectroscopy}
\author{A. Mourachkine}
\address{Universit\'{e} Libre de Bruxelles, CP-232, 
Blvd du Triomphe, B-1050 Brussels, Belgium}

\date{Received 8 September 2000, in final form 1 March 2001;
{\bf Supercond.  Sci. Technol. 14}, 329}
\maketitle

\begin{abstract}
{\bf Abstract.} Tunneling measurements have been carried out on heavily 
underdoped, slightly overdoped and partially Ni-substituted 
Bi$_{2}$Sr$_{2}$CaCu$_{2}$O$_{8+x}$ (Bi2212) single crystals by using
a break-junction technique. We find that in-plane
tunneling spectra below $T_{c}$ are the combination of incoherent 
part from the pseudogap and coherent quasiparticle peaks.
There is a clear correlation between the magnitude of the 
pseudogap and the magnitude of the superconducting gap in Bi2212. 
The analysis of the data suggests that the {\em tunneling} 
\,pseudogap in Bi2212 is predominantly a charge-density-wave gap on
dynamical charge stripes. 
The tunneling characteristics corresponding to the quasiparticle peaks are 
in excellent agreement with theoretical predictions made for a quasi-one 
dimensional topological-excitation liquid. 
In addition, the analysis of data measured by different techniques shows 
that the phase coherence along the $c$\,-axis is established at $T_{c}$ due 
to spin fluctuations in local antiferromagnetic domains of CuO$_{2}$ planes.
\end{abstract}

\vspace{5mm}
]

\section{Introduction}
 
Soon after the discovery of superconductivity in cuprates \cite{Muller}, 
it became clear that the concept of the Fermi liquid is not applicable 
to cuprates: the normal state properties of cuprates are markedly 
different from those of conventional metals \cite{Orenstein}.
The pseudogap (PG) which appears in electronic excitation spectra of 
cuprates above $T_{c}$, 
is one of the main features of high-$T_{c}$ superconductors (SCs). 
There is a consensus on doping dependence of the PG in hole-doped 
cuprates: the magnitude of the PG decreases with increase in 
hole concentration \cite{Orenstein,Timusk,Tallon}.
However, there is a clear discrepancy between the phase diagrams 
inferred from transport measurements, on the one hand, and from 
tunneling measurements, on the other hand: transport 
measurements \cite{Timusk,Tallon,Wuts} show that, in the overdoped 
region, the PG is absent above $T_{c}$, at the same time, in tunneling 
measurements \cite{Yurgens,Ekino1,Matsuda,AMour1,Oda}, the PG is 
observed well above $T_{c}$.

The SC characteristics in cuprates have different 
doping dependences: the $T_{c}$ value has approximately the 
parabolic dependence on hole concentration, $p$, with the maximum 
near $p$ = 0.16 \cite{Tallon}, whereas the SC condensation energy 
has the maximum in the overdoped region near $p$ = 0.19 
\cite{Tallon,Shen}.

Recent intrinsic $c$\,-axis tunneling data obtained in 
Bi$_{2}$Sr$_{2}$CaCu$_{2}$O$_{8+x}$ (Bi2212) mesas \cite{Yurgens} 
show that the pseudogap (PG) is the normal-state gap, and the PG
and the superconducting gap (SG) coexist below $T_{c}$. 
High-resolution angle-resolved photoemission
(ARPES) data \cite{Fedorov} obtained in Bi2212 at momentum near the 
(0, $\pi$) also show that the quasiparticle (QP) peak and the gap 
have different origins, and they coexist below $T_{c}$. Thus, the 
PG in Bi2212 arises from either charge-density waves (CDW) or local 
antiferromagnetic (AF) correlations [or spin-density waves (SDW)], 
or from their combination \cite{Mark,Klem,Gabi2,Gabi1}. It was proposed 
that, in order to develop further understanding of ARPES spectra, 
it is necessary to separate the coherent QP peak from the gap 
in ARPES spectra \cite{Shen}.

Tunneling spectroscopy is an unique probe of SC state in that it can, 
in principle, reveal the QP excitation density of states (DOS) 
directly with high energy resolution.
In this paper we present tunneling measurements performed on 
heavily underdoped, slightly overdoped and partially Ni-substituted 
Bi2212 single crystals by using a break-junction technique. We find 
that in-plane tunneling spectra below $T_{c}$ are the 
combination of incoherent part from the PG and coherent QP 
peaks. There is a clear correlation between the magnitude of the 
PG and the magnitude of the SG in Bi2212.
Analysis of the data suggests that the {\em tunneling} PG 
in Bi2212 is predominantly a CDW gap on dynamical charge stripes. 
The tunneling characteristics corresponding to the QP peaks are in 
excellent agreement with theoretical predictions made for a quasi-one 
dimensional (1D) topological-excitation liquid. In addition, analysis 
of data measured by different techniques shows that the phase coherence 
along the $c$\,-axis is established at $T_{c}$ due to spin fluctuations 
in local AF domains of CuO$_{2}$ planes. In the framework of the quasi-1D 
topological-excitation-liquid scenario in Bi2212, the discrepancy 
between the phase diagrams in the overdoped region, inferred from 
transport and tunneling measurements, becomes obvious: they measure 
two different PGs. Other data obtained in cuprates can be naturally 
understood in the framework of such a scenario for the SC in Bi2212. To 
our knowledge, the observation of a quasi-1D topological-excitation 
liquid in Bi2212 is presented in the literature for the first time. 
 
\section{Theory and measured data}

By performing tunneling measurements one can obtain the $I(V)$ and 
$dI/dV(V)$ characteristics. The $dI/dV(V)$ tunneling characteristic 
measured in a SC-insulator-normal metal (SIN) junction corresponds 
directly to the DOS of QP excitations \cite{Wolf}.
In this Section, first, we consider a theoretical $I(V)$ tunneling 
characteristic in a SIN junction and a measured $I(V)$ curve in a 
SC-insulator-SC (SIS) junction of Bi2212.
 
Figure 1(a) shows a theoretical $I(V)$ characteristic in a SIN 
junction (Fig.6 in Ref.\cite{Tinkham}). In the {\em tunneling} 
regime, it is expected that the $I(V)$ curve at high positive (low 
negative) bias, depending on the normal resistance of junction, lies 
somewhat below (above) the normal-state curve [the dash line in 
Fig.1(a)] \cite{Tinkham}. In conventional SCs, this prediction is 
verified by tunneling experiments \cite{Wolf}. However, we find that, 
in cuprates, the prediction is violated. Figure 1(b) shows the $I(V)$ 
curve measured in a slightly underdoped Bi2212 single crystal 
with $T_c$ = 83 K (Fig.1 in Ref.\cite{Miyakawa2}). In Fig.1(b), 
one can see that the $I(V)$ curve at high positive (low negative) bias 
passes not below (above) the straight dash line but far above (below) 
the line. This fact cannot be explained by the d-wave symmetry of the 
order parameter. To our knowledge, this question has been never 
raised in the literature before. This finding is the main motivation 
of the present work.
  
Before we discuss our tunneling data, it is necessary to consider 
theoretical characteristics of a topological soliton and a bound state of 
two solitons. A topological soliton is an extremely stable nonlinear 
excitation which can be moving or entirely static 
\cite{french,german,Davydov1,Davydov2,Mostovoy}. The topological 
soliton is often called a kink. A kink is a domain wall between two 
degenerate ground states. Solitons and kinks have particlelike properties.
Usually, the kinks appear in half-filled (spin-) Peierls 
systems which are characterized by either a CDW or SDW instability 
\cite{french,german,Davydov1,Davydov2,Mostovoy,Maki2,Mele,Su}. 
The CDW and SDW instabilities in 
(spin-) Peierls systems are the consequence of lattice distortion. For 
example, in polyacetylene, the solitons are moving domain 
walls between the two degenerate dimerized phases \cite{Su}. Solitons 
(kinks) may carry a charge (which may be fractional) or spin of 1/2
or both. Figure 2(a) shows the characteristics of a kink. 
Since, in Fig.2, we consider the general solutions, the gap shown 
{\em schematically} in Fig.2 is either a CDW or SDW gap.  So, the kinks 
occupy the midgap states. A kink (soliton) can form a bound 
state with another kink (soliton), which can be moving or stationary. 
Figure 2(b) shows the characteristics of a bound state of two solitons. 

To our knowledge, a soliton SC was for the first time considered in 
Ref.\cite{bisoliton} in order to explain the SC in quasi-1D organic 
conductors. In this model, two electron-solitons form a bisoliton having 
2$e$ charge and zero spin (a singlet state). The QPs surrounded 
by deformation move coherently as a unique entity along stacks of 
organic molecules without resistance. Later, Davydov 
\cite{Davydov1,Davydov2} applied the model of the bisoliton SC to 
cuprates. Recently, other authors proposed that the charge carriers in the 
normal state of cuprates reside on domain walls as a consequence of 
strong short-range electron-electron repulsion \cite{Nayak}.
\begin{figure}[t]
\leftskip-10pt
\epsfxsize=1.0\columnwidth
\centerline{\epsffile{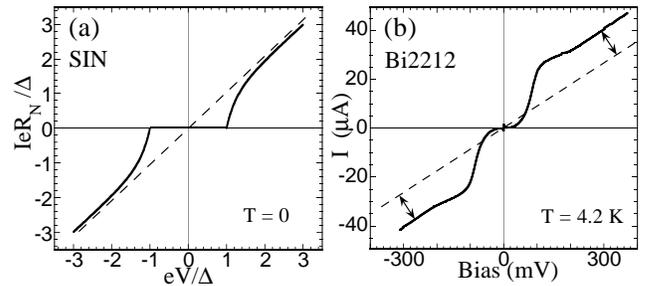}}
\vspace{2mm}
\caption{(a) Theoretical $I(V)$ tunneling characteristic in a SIN 
junction of a SC with the isotropic energy gap \protect\cite{Tinkham}. 
The dash line shows a normal-state curve. 
(b) Measured $I(V)$ curve in a SIS junction of an underdoped Bi2212 
with $T_{c}$ = 83 K \protect\cite{Miyakawa2}. The dash 
line which is parallel to the $I(V)$ curve at high bias is a guide 
to the eye. The arrows show the offset from the dash line.
Note that the dash line in the plot (b) is not the normal-state 
curve. In fact, {\em any} straight line passing through (0,0) will show 
that the behavior of the $I(V)$ curve in the plot (b) deviates 
from the theory \protect\cite{Tinkham}.}
\label{fig1}
\end{figure}
 
\begin{figure}[h]
\leftskip-10pt
\epsfxsize=0.9\columnwidth
\centerline{\epsffile{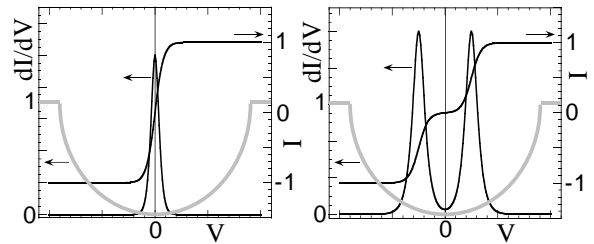}}
\vspace{2mm}
\caption{(a) Normalized $dI/dV(V)$ and $I(V)$ characteristics of a kink. 
(b) Normalized $dI/dV(V)$ and $I(V)$ characteristics of a bound state of 
two solitons \protect\cite{french,Mostovoy}. In both insets: the gap is 
shown 
{\em schematically} in grey; the $I(V)$ curve corresponding to the gap is 
not shown, and the hight of the conductance soliton peaks depends on the 
density of added (removed) electrons \protect\cite{Maki2,Mele}. }
\label{fig2}
\end{figure}
\section{Experiment}
 
Overdoped Bi2212 single crystals and Bi2212 single 
crystals in which Cu is partially substituted for Ni (Ni-Bi2212) 
were grown using a self-flux method. The dimensions of the samples are 
typically of 3$\times$1$\times$0.1 mm$^3$. The chemical composition of
the Bi-2:2:1:2 phase in the overdoped crystals corresponds 
to the formula Bi$_{2}$Sr$_{1.9}$CaCu$_{1.8}$O$_{8+x}$ as measured by
energy  dispersive X-ray fluorescence (EDAX). The crystallographic $a$,
$b$, $c$ values of the overdoped crystals are 
5.41 \AA, 5.50 \AA\,\,and 30.81 \AA, 
respectively. The chemical composition of the  
Bi-2:2:1:2 phase  in the Ni-Bi2212 crystals corresponds to the 
formula Bi$_{2}$Sr$_{1.95}$Ca$_{0.95}$(CuNi)$_{2.05}$O$_{8+x}$ 
as measured by EDAX. The Ni content with respect to 
Cu is approximately 1.5\%. The $T_{c}$ value was determined 
by four-contact method. The transition width is less than 1 K in
the overdoped Bi2212 crystals, and a few degrees in the Ni-Bi2212 
single crystals. The underdoped samples were obtained from the 
overdoped single crystals by annealing them in vacuum.

The results presented here are obtained in three single crystals:
one underdoped, one overdoped and one Ni-Bi2212, having the 
$T_{c}$ values of 51, 88 and 75, respectively. The hole 
concentration in each of them estimated from empirical 
relation $T_{c}/T_{c,max} = 1 - 82.6(p - 0.16)^{2}$ \cite{Tallon}
is equal to 0.085, 0.19 and 0.2, respectively. 
We use $T_{c,max}$ = 95, 95 and 87 K, respectively.
The maximum $T_{c}$ value for the Ni-Bi2212 single crystals is 
estimated from $dT_{c}/dn_{Ni} \simeq$ -5 K/at.\%, where 
$n_{Ni}$ is the Ni content with respect to Cu \cite{Ni}.
\begin{figure}[t]
\leftskip-10pt
\epsfxsize=1.0\columnwidth
\centerline{\epsffile{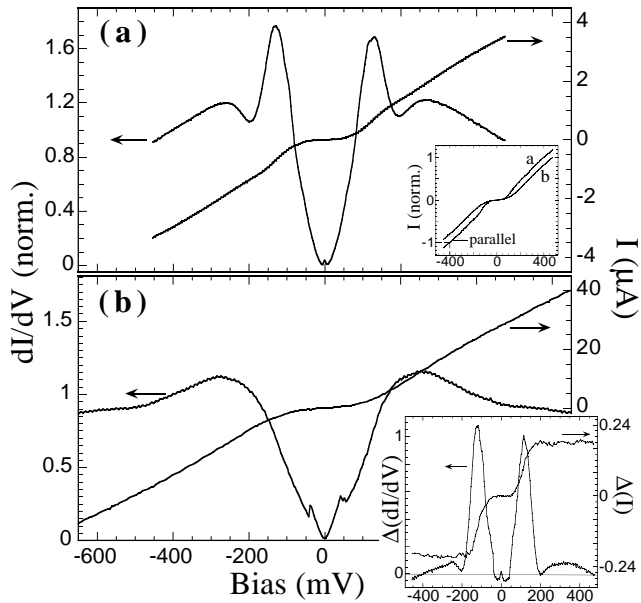}}
\vspace{2mm}
\caption{SIS $dI/dV(V)$ and $I(V)$ measured at 14 K 
within the same underdoped Bi2212 single crystal with $T_{c}$ = 51 K.
The $dI/dV(V)$ in both plots are normalized at -400 mV. Inset 
in the plot (b): the differences $(dI/dV)_{a}- (dI/dV)_{b}$ 
and $I_{a,norm}- I_{b,norm}$. The inset in the plot (a) shows how 
$I_{a}$ and $I_{b}$ were normalized: $I_{a}$ is normalized at 
-400 mV, and $I_{b}$ is adjusted to be parallel at high bias to 
$I_{a}$ (such procedure is equivalent to the normalization at 
$\pm\infty$).}
\label{fig3}
\end{figure}
Initially, the tunneling measurements in the Ni-Bi2212 single crystals 
were motivated by the presence of magnetic impurities (Ni) in Bi2212. 
However, we use the data obtained in the Ni-Bi2212 as data in the 
overdoped region ($p$ = 0.2) without emphasizing the presence of Ni.

Experimental details of our break-junction setup can be found 
elsewhere \cite{Hancotte}. In short, many break-junctions were 
prepared by gluing a sample with epoxy on a flexible insulating 
substrate, and then were broken in the $ab$\,-plane by bending the 
substrate with a differential screw at low temperature in a He 
ambient. The electrical contacts (typically with 
the resistance of a few Ohms) were made by attaching gold wires 
to a crystal with silver paint. The $I(V)$ and $dI/dV(V)$ 
tunneling characteristics were determined by the four-terminal 
method by using a standard lock-in modulation technique.
At low (constant) temperature, in one junction, we usually obtain a 
few tunneling spectra by changing the distance between broken parts of 
a crystal, going back and forth {\em etc}., and, every time, the tunneling
occurs most likely in different places.

In addition to SIS-junction measurements, tunneling tests have been
carried out in the overdoped Bi2212 single crystals by forming SIN 
junctions. Pt-Ir wires sharpened mechanically are used as normal 
tips.

The magnitude of the SG can, in fact, be derived directly from the 
tunneling spectrum. However, in the absence of a generally accepted 
model for the gap function and the DOS in cuprates, we calculate the 
gap magnitude 2$\Delta$ as a half spacing between the coherent QP 
peaks in SIS conductance tunneling characteristics.

\section{Results}
\subsection{The underdoped region}

Figure 3(a) shows the SIS $dI/dV(V)$ and $I(V)$ 
curves obtained in the underdoped Bi2212 single crystal,  
which look like usual tunneling spectra in Bi2212 \cite{Miyakawa}. 
In Fig.3(a), the Josephson $I_{c}R_{n}$ product is estimated to be 
13.4 mV. The gap magnitude, $\Delta_{sc}$ = 64 meV, is in good 
agreement with other tunneling measurements \cite{Miyakawa}. The 
$dI/dV(V)$ and $I(V)$ curves shown in Fig.3(b) are 
obtained within the {\em same} underdoped single crystal as those in 
Fig.3(a). In Fig.3(b), the wide humps resemble the humps in the 
$dI/dV(V)$ shown in Fig.3(a). It is suggestive that the spectra in 
Fig.3(b) correspond to the PG. The SC in Bi2212 is weak in the heavily 
underdoped region ($p <$ 0.1) \cite{Tallon,Shen}. For example, in 
YBa$_{2}$Cu$_{3}$O$_{6+x}$ (YBCO), modest magnetic fields suppress 
the SC significantly in the heavily underdoped 
region \cite{Dai}. This may explain why it is possible 
to observe separately the PG in the heavily underdoped Bi2212 by 
taking into account that tunneling spectroscopy probes the local DOS. 
The absence of the Josephson current in the spectra shown in Fig.3(b) 
indicates that the humps in the conductance are incoherent. 
The differences between the spectra 
shown in Figs 3(a) and 3(b) are presented in the inset of Fig.3(b), 
which correspond to a "pure SG". Some parts of the $dI/dV(V)$ in the 
inset of Fig.3(b) are slightly below zero because the spectra shown 
in Figs 3(a) and 3(b) are not taken under the exact same conditions. 
The small humps in $dI/dV(V)$ shown in the inset of Fig.3(b) are 
discussed below. The $dI/dV(V)$ and $I(V)$ spectra in the inset of Fig.3(b) 
resemble the characteristics of a bound state of two solitons, shown 
in Fig.2(b). Thus, we find that the conductance in Fig.3(a) consists of 
two contributions: the incoherent humps which correspond to the PG (which 
is a normal-state gap \cite{Yurgens}), and the coherent QP peaks from 
condensed solitonlike excitations. In Fig.3(b), one can see that the PG is 
anisotropic.

We discuss now the PG. Quasi-1D topological excitations reside 
on quasi-1D "objects". Recently, quasi-1D charge stripes separated by
2D AF domains have been discovered in some 
cuprates \cite{Tranquada,AMour2}. The presence of solitonlike excitations 
on stripes implies that the stripes are {\em insulating}. In other words, 
there is a charge gap along stripes, which is most likely a CDW gap Then, 
the {\em tunneling} PG is predominantly a CDW gap. The same conclusion 
can be independently obtained from the analysis of the data: The 
magnitude of PG, $\Delta_{ps}$ = 130 meV, shown in Fig.3(b), is 
too large to be explained by the development of local 
AF correlations, since the value of the 
superexchange energy, $J \simeq$ 120 meV \cite{Bourge}, is not large 
enough to fit the data. Moreover, by taking into account that the PG 
increases with decrease in hole concentration, at lower doping 
($p <$ 0.085), the magnitude of the PG will become larger than 130 
meV. Thus, one can conclude that the PG is too large to be the spin gap 
due to local AF correlations. For the second time, independently, we 
find that the {\em tunneling} PG is most likely a CDW gap. 

It is possible that the subgap in the $dI/dV(V)$ shown in Fig.3(b) 
is due to the stripe excitations which were damped at higher 
bias by tunneling electrons. The SC is weak in the underdoped 
region \cite{Tallon,Shen,Dai}.

\subsection{The overdoped region}

Figures 4(a) and 4(b) show the SIS $dI/dV(V)$ and $I(V)$ 
curves obtained within the {\em same} overdoped Bi2212 single 
crystal. In Figs 4(a) and 4(b), the Josephson $I_{c}R_{n}$ product is 
estimated to be 6 and 7.5 mV, respectively. The $dI/dV(V)$ 
and $I(V)$ characteristics shown in Fig.4(a) look like 
usual tunneling spectra in Bi2212 \cite{Miyakawa}. The $dI/dV(V)$ 
and $I(V)$ curves in Fig.4(b) resemble the spectra shown in the inset 
of Fig.3(b), and the characteristics of a bound state of two solitons
in Fig.2(b). This means that, the contribution from the PG in the 
tunneling spectra shown in Fig.4(b) is small in comparison with the 
contribution from the QP peaks, at least, at low bias. This is most 
likely due to the fact that, in slightly overdoped cuprates, the SC is 
the strongest, and the "strength" of the PG is weak \cite{Tallon,Shen}. 
At high bias, the contribution from the PG will be always predominant, 
even if the PG is weak. 

In order to show that we observe not a SIS-junction effect 
but an intrinsic effect, we performed measurements in the
overdoped Bi2212 single crystals by SIN junctions.
The inset of Fig.4(a) shows the SIN $dI/dV(V)$ and $I(V)$ 
curves obtained in an overdoped Bi2212. In Fig.4(a), one can
see that, basically, there is no difference between the $I(V)$
characteristics measured in SIS and SIN junctions [see the dash lines
in Fig.4(a) and in the inset of Fig.4(a)].

The inset of Fig.4(b) shows the PG in the slightly overdoped Bi2212 
at 122 K, which looks like the {\em typical} CDW gap in quasi-1D 
conductor NbSe$_{3}$ \cite{He,Ekino2}. The value of 122 K is chosen 
not by accident. Two independent tunneling studies show that, in slightly 
overdoped Bi2212, the onset of SC occurs at 110--116 K \cite{Ekino1,AMour1}. 
Thus, the value of 122 K is above the onset of SC in 
slightly overdoped Bi2212. In the inset of Fig.4(b), one can see 
that the PG is anisotropic. The shape of the PG at low bias indicates that
it has most likely an anisotropic s-wave rather than the d-wave shape.
\begin{figure}[t]
\leftskip-10pt
\epsfxsize=1.0\columnwidth
\centerline{\epsffile{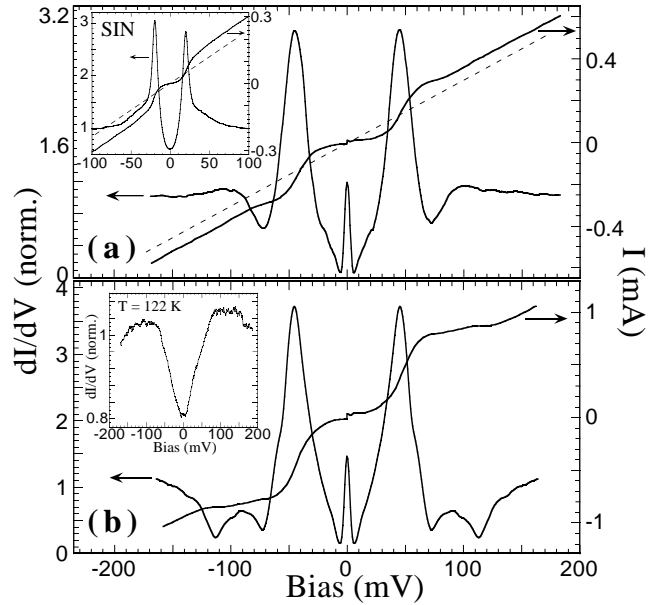}}
\vspace{2mm}
\caption{SIS $dI/dV(V)$ and $I(V)$
measured at 14 K within the same overdoped Bi2212 single crystal 
with $T_{c}$ = 88 K. Inset in the plot (a): SIN $dI/dV(V)$ and 
$I(V)$ measured at 9 K in an overdoped Bi2212 with $T_{c}$ = 87.5 K 
(same axis parameters as main plot). The dash lines in the plot 
(a) and in the inset, which are parallel at high bias to the $I(V)$ 
curves, are guides to the eye. Inset in the plot (b): the PG measured 
at 122 K.}
\label{fig4}
\end{figure} 

We discuss now the tunneling data measured in Ni-Bi2212
single crystals which are also overdoped in oxygen. Figures 5(a) and 
5(b) show the SIS $dI/dV(V)$ and $I(V)$ curves obtained 
within the {\em same} Ni-Bi2212 single crystal. In Figs 5(a) and 5(b), 
the Josephson $I_{c}R_{n}$ product is estimated to be 3.1 and 24.5 
mV, respectively. The SIS $dI/dV(V)$ and $I(V)$ 
curves shown in Fig.5(a) look like usual tunneling spectra in 
Bi2212 \cite{Miyakawa}. The $dI/dV(V)$ and $I(V)$ curves in 
Fig.5(b) resemble the spectra in the inset of Fig.3(b) and the 
characteristics of a bound state of two solitons shown in Fig.2(b). The 
effect of the absence of the contribution from the PG in the tunneling 
spectra shown in Fig.5(b) is even stronger than that in Fig.4(b). It is 
important to emphasize that the spectra measured in underdoped Bi2212, 
shown in in the inset of Fig.3(b), and the spectra measured in overdoped 
Bi2212 and Ni-Bi2212, shown in Figs 4(b) and 5(b), are similar. So, the  
data obtained in underdoped and overdoped Bi2212 are consistent with 
each other.
However, in the overdoped region, the "strengths" of the PG and SG 
are reversed in comparison with those in the underdoped region. In the 
overdoped region, the SC is the strongest, and the PG is 
weak \cite{Tallon,Shen}. In Fig.5(a), one can see that the humps in the 
conductance measured in Ni-doped Bi2212 are weaker than those in 
Figs 3(a) and 4(a). This means that Ni destroys the PG (CDW gap), so 
stripes become more metallic. This is true from another point of view: 
many of our attempts to measure the PG above $T_{c}$ in Ni-Bi2212 
have failed.
\begin{figure}[t]
\leftskip-10pt
\epsfxsize=1.0\columnwidth
\centerline{\epsffile{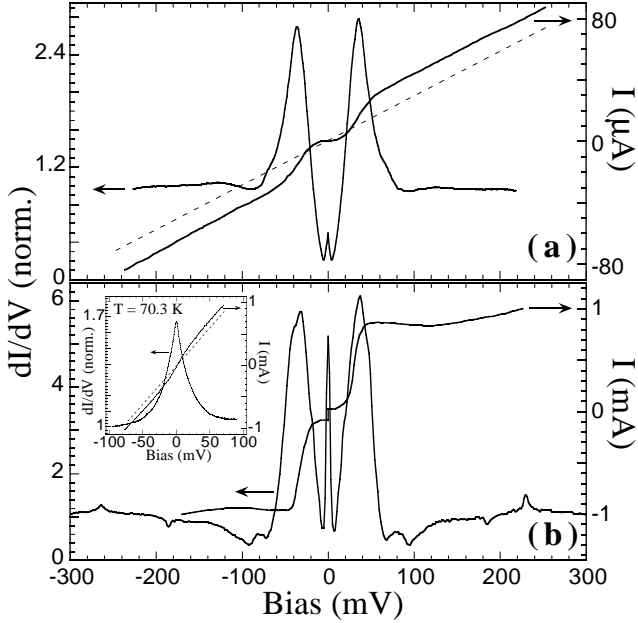}}
\vspace{2mm}
\caption{SIS $dI/dV(V)$ and $I(V)$ measured at 15 K within the same 
Ni-Bi2212 single crystal with $T_{c}$ = 75 K. Inset in the plot (b): 
$dI/dV(V)$ and $I(V)$ characteristics of the main plot at 70.3 K. The 
dash lines which are parallel to the $I(V)$ curves at high bias are 
guides to the eye.}
\label{fig5}
\end{figure}

Figure 6 displays temperature dependence of the conductance
shown in Fig.5(b). In Fig.6, the QP peaks disappear {\em below} 
$T_{c}$ = 75 K. This is due to the fact that tunneling spectroscopy 
probes the local DOS; the $T_{c}$ is the macroscopic 
characteristic. Thus, in Fig.6, $T_{c,local} \simeq$ 70 K.
Apparently, the data shown in Fig.6 are measured near an impurity 
({\em i.e.}\,Ni). It seems that the small humps which appear in 
the $dI/dV(V)$ shown in Figs 4(b) and 5(b) at bias {\em twice} as large 
as bias of the QP peaks,  relate to the QP peaks and not to the PG. 
In Fig.6, their temperature dependence resembles the temperature 
dependence of a SG. In the inset of Fig.3(b), a similar hump is present in 
the conductance at negative bias. These humps can be well understood in 
terms of a {\it nanopteron} soliton \cite{french} which is discussed below.
\begin{figure}[t]
\leftskip-10pt
\epsfxsize=0.7\columnwidth 
\centerline{\epsffile{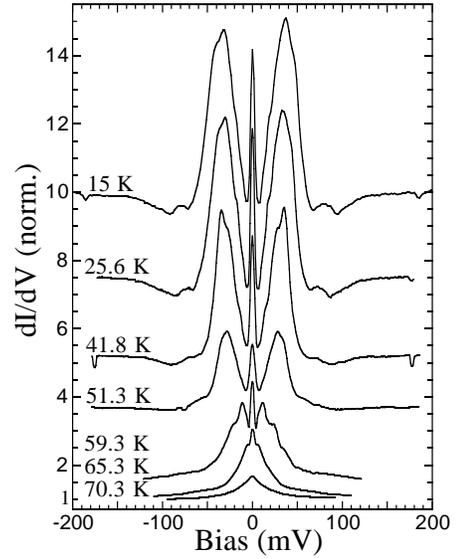}}
\vspace{2mm}
\caption{Temperature dependence of the conductance peaks 
shown in Fig.5(b). The conductance scale corresponds to the 70.3 K 
spectrum, the other spectra are offset vertically for clarity.}
\label{fig6}
\end{figure}

The $dI/dV(V)$ and $I(V)$ characteristics at 70.3 K in the Ni-Bi2212 
are shown separately in the inset of Fig.5(b). If we subtract the 
linear contribution in the $I(V)$ curve shown in the inset
of Fig.5(b) (see the dash line), then the $dI/dV(V)$ and $I(V)$  
spectra in the inset of Fig.5(b) resemble the characteristics of the kink 
shown in Fig.2(a).
\vspace{1mm}
\begin{figure}[h]
\leftskip-10pt
\epsfxsize=0.8\columnwidth
\centerline{\epsffile{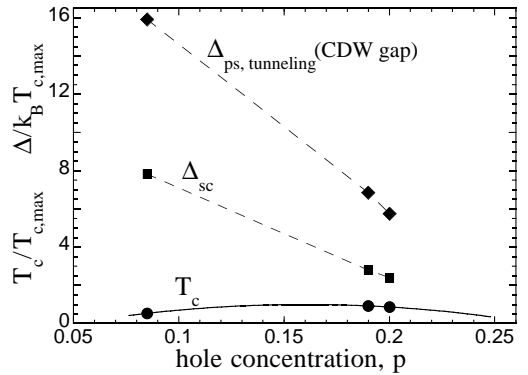}}
\vspace{2mm}
\caption{Low temperature phase diagram of Bi2212 based on the
present data: $\Delta$$_{sc}$ (squares), $\Delta$$_{ps}$ (diamonds), 
and $T_{c}$ (dots) The solid line corresponds
to $T_{c}/T_{c,max} = 1 - 82.6(p - 0.16)^{2}$ \protect\cite{Tallon}. 
The dash lines are guides to the eye.}
\label{fig7}
\end{figure}

\subsection{The phase diagram}

Figure 7 depicts the phase diagram of Bi2212 based on the present 
data. The magnitude of the SG is in good agreement with other 
tunneling measurements \cite{Miyakawa}. In Fig.7, one can see that 
$\Delta$$_{sc}$ and $T_{c}$ do not correlate with each other. 
As mentioned above, the {\em tunneling} PG shown in Fig.7 is different 
from the PG inferred from transport 
measurements \cite{Timusk,Tallon,Wuts}.

\section{The measured data and theory}

We compare now the tunneling data with theory. The characteristics of 
a kink and a bound state of two solitons are described by hyperbolic 
functions \cite{french,german,Davydov1,Davydov2,Mostovoy}. Since the 
conductance peaks of a bound state of two solitons, which are shown in  
Fig.2(b), look very similar to the conductance peaks not only of 
high-$T_{c}$ SCs but also of low-$T_{c}$ SCs (not the background), 
we rely here exclusively on the $I(V)$ characteristics which 
are {\em conceptually} different for the two models based on the Fermi 
liquid, in the first model, and on a quasi-1D topological-excitation liquid, 
in the second model. 

\subsection{$I(V)$ characteristics}

Figure 8(a) shows the measured $I(V)$ curve from the inset of Fig.3(b). 
In Fig.8, for simplicity, we analyze the data only at positive bias. As shown 
in Fig.8(a), the data from the inset of Fig.3(b) can be fitted very well by the 
hyperbolic function
$f(V)$ = $A$$\times$$(tanh[(eV$ - 2$\Delta$$)/eV_{0}] 
+ tanh[(eV$ + 2$\Delta$$)/eV_{0}])$, where $e$ is the electron 
charge; $V$ is the bias; $\Delta$ is the maximum SC energy gap, and 
$A$ and $V_{0}$ are the constants. In Fig.8(a), we also present the 
measured $I(V)$ curve of the PG  from Fig.3(b). We find that any 
tunneling $I(V)$ curve in Bi2212 can be resolved into the two 
components shown in Fig.8(a). Thus, the $I(V)$ characteristics in 
Bi2212 (as well as the $dI/dV(V)$ characteristics) consist of two 
contributions: from the quasi-1D topological excitations and from the 
PG which is the normal-state gap \cite{Yurgens}. The "usual" $I(V)$ 
and $dI/dV(V)$ spectra in Bi2212 show the presence of both components 
[see Figs 3(a), 4(a) and 5(a)]. The absence [see Fig.3(b)] or weak contribution 
of one component [see Figs 4(b) and 5(b)] in tunneling spectra make the 
appearance of the spectra "unusual".

First, we analyze the tunneling data from Ref.\cite{Miyakawa2}: 
Figure 8(b) shows the data from Fig.1(b), the $f(V)$ fit, and their 
difference. Figure 8(c) depicts the two components in the 
$I(V)$ curve from Fig.4(b). As shown in Fig.8(d), 
the contribution from the QP peaks in the SIN $I(V)$ 
from the inset of Fig.4(a) seems to be weaker than that in SIS 
junctions. As seen in Fig.8, all plots are similar.
To fit the SIN $I(V)$, we use the same $f(V)$ function by 
substituting 2$\Delta$ for $\Delta$. In Figs 8(b)--8(d), the amplitude, 
$A$, of the $f(V)$ fit can be changed, this only affects the scale but 
not the shape of the differences which correspond to the PG. The 
decomposition of the $I(V)$ curves in Figs 8(b)--8(d) into 
the two components explains why $I(V)$ tunneling characteristics in 
cuprates do not obey the theoretical predictions based on the 
Fermi-liquid model \cite{Tinkham}. The $I(V)$ curves in 
Figs 4(a) and 5(a) and in Refs \cite{Miyakawa2} and 
\cite{Miyakawa} can be resolved into the two components in the 
same manner. The data in Fig.3 are self-consistent.
\begin{figure}[t]
\leftskip-10pt
\epsfxsize=1.0\columnwidth
\centerline{\epsffile{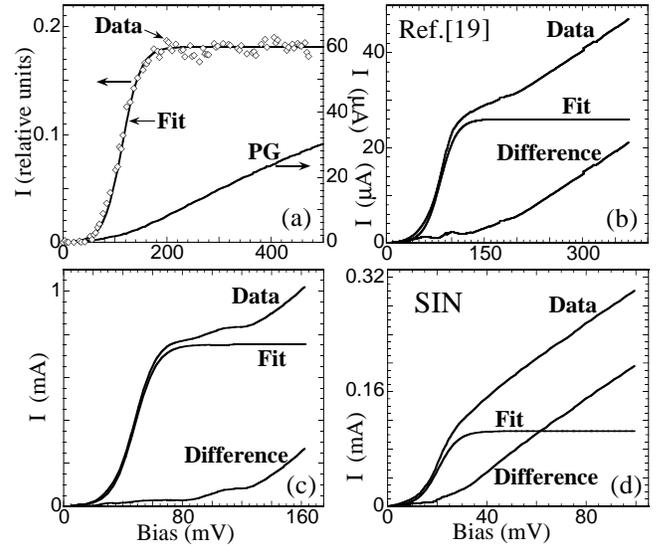}}
\vspace{2mm}
\caption{Measured $I(V)$ curves and the $f(V)$ fit (see text)
in (a)-(b) underdoped and (c)-(d) overdoped Bi2212: 
(a) The data (diamonds) from the inset of Fig.3(b); the measured 
$I(V)$ of the PG from Fig.3(b), and the $f(V)$ fit. 
(b) The data from Ref.\protect\cite{Miyakawa2}, shown in Fig.1(b); 
the $f(V)$ fit, and their difference. (c) The $I(V)$ from Fig.4(b); 
the $f(V)$ fit, and their difference. (d) The SIN $I(V)$ 
from the inset of Fig.4(a); the $f(V)$ fit 
[2$\Delta \rightarrow \Delta$ in $f(V)$], and their difference. 
In the plots (b) and (c), the Josephson currents are removed.}
\label{fig8}
\end{figure}
\begin{figure}[h]
\leftskip-10pt
\epsfxsize=1.0\columnwidth
\centerline{\epsffile{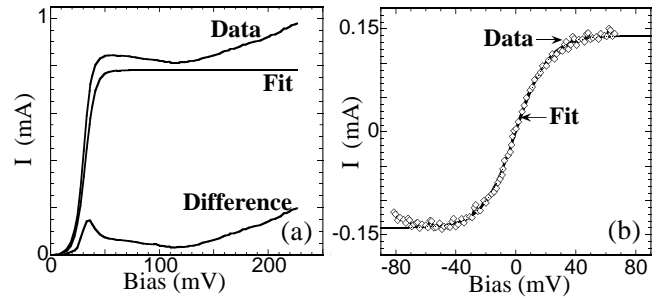}}
\vspace{2mm}
\caption{Measured $I(V)$ curves and the $f(V)$ and $f_{n}(V)$ fits 
(see text) in Ni-Bi2212: (a) The $I(V)$ from Fig.5(b); the $f(V)$ 
fit, and their difference. (b) The $I(V)$ from the inset of 
Fig.5(b) (diamonds) (the linear background is subtracted) and the 
$f_{n}(V)$ fit. In the plot (a), the Josephson current is removed.}
\label{fig9}
\end{figure}

However, we find that the $I(V)$ curve shown in Fig.5(b)
consists of three components. Figure 9(a) shows the $I(V)$
from Fig.5(b), the $f(V)$ fit and their difference.
In Fig.9(a), it is clear that the difference consists of two 
contributions: the contribution at high bias is predominantly
from the PG, similar to those in Fig.8, and the
contribution at low bias with the peak at 35 mV. By analyzing the 
data we show below that this peak corresponds to electron 
tunneling assisted by spin excitations \cite{Tsui}. In other words, 
this peak corresponds to the magnetic resonance peak observed in 
inelastic neutron scattering experiments \cite{Mignod}.

Figure 9(b) shows the $I(V)$ curve from the inset of 
Fig.5(b), where the linear contribution [see the dash line in the 
inset of Fig.5(b)] is subtracted. As mentioned above, the $I(V)$
curve in Fig.9(b) is very similar to the $I(V)$ 
characteristic of a kink, shown in Fig.2(a). This means 
that, at this temperature, the quasi-1D topological excitations are 
already decoupled. We use the hyperbolic function 
$f_{n}(V)$ = $A_{1}$$\times$$tanh(V/V_{0})$ 
to fit the data in Fig.9(b), which gives a good interpolation. 
Thus, there is good agreement between the data below $T_{c}$ 
and above ({\em local}) $T_{c}$.

In {\em conventional} SCs, tunneling conductances measured in SIN and SIS 
junctions contain different information. If a SIN conductance corresponds 
directly to the SC DOS, a SIS conductance is the convolution of the DOS with 
itself \cite{Wolf}. In Fig.8, we use the same function to fit $I(V)$ 
characteristics measured in SIS {\em and} SIN junctions. Physically, it is 
incorrect. However, the {\em main point} of the fit is that the asymptotics 
of $I(V)$ characteristics corresponding to QP peaks, measured either in SIS 
or SIN junctions, are ''flat'' (constant). The ''flat'' asymptotics in 
tunneling spectra are a hallmark of the presence of one dimensionality 
in the system. Such asymptotics never appear in 2D or 3D systems.

So, we conclude that, in Bi2212, the $I(V)$ characteristics 
corresponding to the QP peaks {\em definitely} disagree with 
Blonder-Tinkham-Klapwijk's predictions based on
the Fermi-liquid model \cite{Tinkham}, and are in good agreement with 
the theoretical predictions made for a quasi-1D topological-excitation 
liquid \cite{french,german,Davydov1,Davydov2,Mostovoy}.

\subsection{$dI/dV(V)$ characteristics}

As emphasized above, we do not rely on the $dI/dV(V)$ data in order 
to draw any conclusion about the origin of QPs in Bi2212. However, 
the analysis of the data would be not complete, if we did not consider 
the conductance spectra. The coherent QP peaks in the 
SIS tunneling spectra can be fitted very well by the derivative
$[f(V)]'$ = $A_{2}$$\times$$((sech[(eV$ - 2$\Delta$$)/eV_{0}])^{2} + 
(sech[(eV$ + 2$\Delta$$)/eV_{0}])^{2})$, as shown in Fig.10.
The fit is applicable only to the coherent QP peaks, and not to the 
humps at high bias, which correspond to the PG, and neither to the 
Josephson current. The $dI/dV(V)$ curves in Figs 4(a) and 
5(a) and in Refs \cite{Miyakawa2} and \cite{Miyakawa} can be 
fitted in the same manner. For example, the conductance in an overdoped 
Bi2212 single crystal, shown in Fig.2 of Ref.\cite{Miyakawa2}, 
clearly is the combination of QP peaks and humps.
\begin{figure}[t]
\leftskip-10pt
\epsfxsize=1.0\columnwidth
\centerline{\epsffile{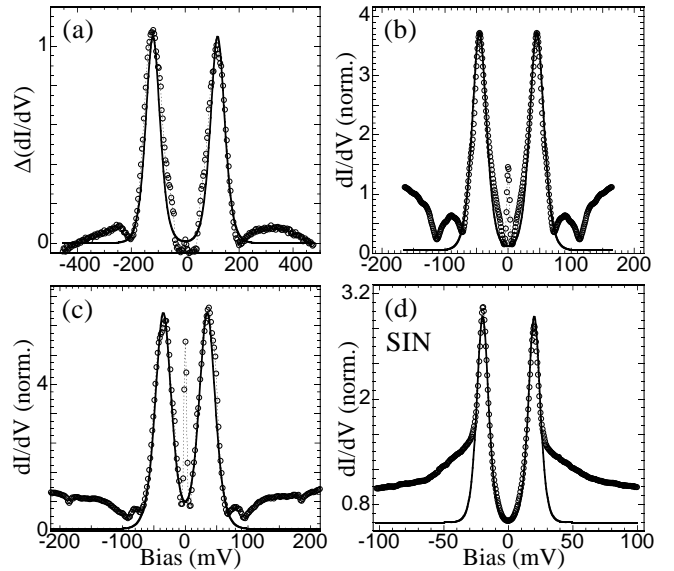}}
\vspace{2mm}
\caption{Measured $dI/dV(V)$ curves (circles) and the $[f(V)]'$ fit 
(see text): (a) $dI/dV(V)$ from the inset of Fig.3(b); 
(b) $dI/dV(V)$ from Fig.4(b); (c) $dI/dV(V)$ from Fig.5(b); 
(d) SIN $dI/dV(V)$ from the inset of Fig.4(a) 
(2$\Delta \rightarrow \Delta$ in $[f(V)]'$).}
\label{fig10}
\end{figure}

In Fig.10(c), the coherent QP peaks contain also
a contribution from electron tunneling due to spin excitations 
[see Fig.9(a)]. As seen in Figs 5(b) and 6, the tops of the QP peaks
have a composite structure which we attribute to the presence of the 
contribution from electron tunneling due to spin excitations: 
This contribution is literally developed on top of the QP peaks.
In Fig.10(d), in order to fit the SIN $dI/dV(V)$ at low bias, we had 
to shift up the $[f(V)]'$ curve. 

The conductance shown in the inset of 
Fig.5(b) can be fitted by $[f_{n}(V)]'$ = $A_{3}$$\times[sech(V/V_{0})]^{2}$.

\subsection{Small humps}

The small humps which appear in the $dI/dV(V)$ at bias twice 
larger than bias of the QP peaks, shown in Figs 4(b) and 5(b) as well as in 
Figs 10(b) and 10(c), can be understood in terms of a nanopteron soliton 
\cite{french}. A nanopteron is a bound state resulting from the nonlinear
interaction between the soliton (or kink) and the periodic wave (see 
Fig.6.27(c) in Ref.\cite{french}). Instead of flat asymptotics, the 
characteristics of a nanopteron have oscillations. A bound state of two 
solitons can also 
exist in resonance with small amplitude linear waves (see Fig.1 in 
Ref.\cite{Buryak}). Since, in Figs 4(b) and 5(b), the predominant 
contribution in the conductances at high bias arises from the PG, it is 
impossible to follow the oscillations in the conductances at high bias. 
Fortunately, the oscillations can be also seen in the $I(V)$ characteristics 
shown in Figs 8(a) and 8(c). In Fig.8(a), the oscillations can be observed up 
to 480 mV. In Fig.8(c), only one hump is present in the $I(V)$ characteristic 
at about 105 mV. The next question is to understand what periodic waves 
in Bi2212 can interact with the bound states of two solitons and cause 
these oscillations? 
Phonons are the primary candidate on this role. It is also possible that
magnonlike excitations (spin 1, charge 0) which cause the appearance of 
the magnetic resonance peak in neutron spectra \cite{Mignod} correspond 
to these linear waves, which are in resonance with bound states of 
topological solitons.

It is well known that $dI/dV(V)$ characteristics measured in SIS junctions 
often have subgap structures [see, for example, Fig.4(a)]. These subgap 
structures can be also understood in terms of a nanopteron soliton. They 
most likely have the same origin as the oscillations at high bias, {\em i.e.} 
due to the nonlinear interaction between a bound state of two solitons and 
the periodic wave (see Fig.4 in Ref.\cite{Buryak}).

\section{Stripes and topological excitations} 

Here we discuss the stripes in cuprates and stripes excitations.

Analysis of the data obtained in inelastic neutron scattering 
experiments suggests that the stripes in cuprates are of the 2$k_{F}$ type 
\cite{Bosch}. The chains in YBCO also have the 2$k_{F}$ charge modulation,
as observed by tunneling spectroscopy 
\cite{Alex}. The 2$k_{F}$ ordering pattern on a stripe is shown 
schematically in Fig.11(a). Recent simulations of ARPES data in Bi2212 
show that the stripes in Bi2212 are most likely site-centered 
\cite{Zacher}. So we consider only the 2$k_{F}$ site-centered 
stripes.

The stripes in cuprates are dynamical \cite{Tranquada}. Recently, 
the dynamics of insulating charge stripes was studied by Zaanen and 
co-workers \cite{Bosch,Eskes,Zaanen}. Figures 11(b) and 11(c) show 
a soliton and a kink on 2$k_{F}$ stripes, respectively. In the 2$k_{F}$
charge modulation along stripes, a kink  (soliton) carries an $e$ charge,
consequently, a bi-soliton has 2$e$  charge. Apparently, charge carriers
in nanotubes are also topological  solitons \cite{nanotubes}.

The soliton SC scenario, probably, is realized in quasi-1D organic metals 
\cite{organic}, 
ladders \cite{Nagata}, and on CuO chains in YBCO which have a CDW ground 
state \cite{Grevin}. The latter fact can be naturally understood since the 
stripes in CuO$_{2}$ planes have also the CDW ground state. Recently, midgap 
solitons have been observed on CuO chains in YBCO by tunneling spectroscopy 
\cite{deLozanne}. The spectrum averaged along a CuO chain shows that there 
is a weak bound state of solitons inside the CDW gap. The magnitude of the 
{\em induced} SG of the bound state of solitons is about 6 meV.
\begin{figure}[t]
\leftskip-10pt
\epsfxsize=0.8\columnwidth
\centerline{\epsffile{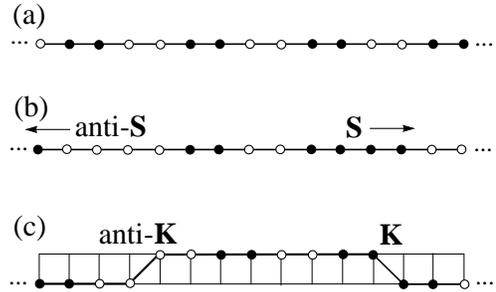}}
\vspace{2mm}
\caption{2$k_{F}$ half-filled stripes: (a) reference state; 
(b) soliton (S) and antisoliton, and (c) kink (K) and antikink.
The full/empty circle denotes the presence/absence of 
the hole.}
\label{fig11}
\end{figure}

Charged solitons repel each other. In conventional SCs, two electrons which 
form a Cooper pair also repel each other. The occurrence of an attractive 
potential between electrons is central to the SC state. So, charged solitons 
will couple with each other if there is an attractive potential between them.

The next question is to understand how is the long-range phase coherence 
established? In CuO$_{2}$ planes, the phase coherence can occur, for example, 
due to the Josephson coupling between SC stripes. Indeed, the so-called 
Yamada plot suggests that $T_{c}$ increases if and only if stripes move closer 
together \cite{Yamada}. Since the SC in CuO$_{2}$ planes is 
{\em fully} 2D because the coupled excitations reside on stripes, the phase 
coherence along the $c$\,-axis is most likely established due to a different 
mechanism.

\section{$c$-axis phase coherence and spin fluctuations in cuprates}

There is a consensus that the SC in cuprates is two-dimensional (2D). 
It is widely believed that the long-range phase coherence occurs at $T_{c}$ 
due to the Josephson coupling between SC CuO$_{2}$ (bi-, tri-, ...)layers. 
Recent measurements of the in-plane ($\rho$$_{ab}$) and out-of-plane 
($\rho$$_{c}$) resistivities in Tl$_{2}$Ba$_{2}$CaCu$_{2}$O$_{8}$ as a 
function of applied pressure show that $\rho$$_{c}(T)$ shifts smoothly down 
with increase of pressure, however, $T_{c}$ first increases and then 
{\em decreases} \cite{Tl2212}. This result can not be explained 
by the interlayer Josephson-coupling mechanism. The authors conclude 
\cite {Tl2212}: ``Any model that associates high-$T_{c}$ with the interplane 
Josephson coupling should therefore be revisited.'' 

Here we analyze data obtained in Andreev reflection, inelastic neutron 
scattering (INS), microwave, muon spin relaxation ($\mu$SR), tunneling and 
resistivity measurements performed on different cuprates, mainly, on YBCO, 
Bi2212 and La$_{2-x}$Sr$_{x}$CuO$_{4+y}$ (LSCO). 
Analysis of the data shows that the long-range phase coherence in the 
cuprates intimately relates to AF interactions along the $c$ axis. 
We also analyze data measured in heavy fermions UPt$_{3}$, 
UPd$_{2}$Al$_{3}$ and CeIrIn$_{5}$, and in some layered non-SC compounds 
with ferromagnetic (FM) correlations.

\subsection{Introduction} 

SC and magnetism were earlier considered as mutually 
exclusive phenomena. Recent research revealed a rich variety of 
extraordinary SC and magnetic states and phenomena in novel materials 
that are due to the interaction between SC and magnetism \cite{Maple}. 
The coexistence of SC and long-range AF order was first discovered in 
$R$Mo$_{6}$Se$_{8}$ ($R$ = Gd, Tb and Er), $R$Rh$_{4}$B$_{4}$ ($R$ = Nd, 
Sm and Tm), and $R$Mo$_{6}$S$_{8}$ ($R$ = Gd, Tb, Dy and Er) \cite{Maple}. 
Later, coexistence of SC and AF order was found in U-based heavy fermions 
(UPt$_{3}$, URu$_{2}$Si$_{2}$, UNi$_{2}$Al$_{3}$, UPd$_{2}$Al$_{3}$, 
U$_{6}$Co, and U$_{6}$Fe), in heavy fermions $R$Rh$_{2}$Si$_{2}$ ($R$ = 
La and Y), Cr$_{1-x}$Re$_{x}$, CeRu$_{2}$, in borocarbides 
$R$Ni$_{2}$B$_{2}$C ($R$ = Tm, Er, Ho, Dy), in organic SCs 
\cite{Gabi1,Maple} and in the new heavy fermion 
CeRh$_{1-x}$Ir$_{x}$In$_{5}$ \cite{LosAmos}. In 
CeRh$_{0.5}$Ir$_{0.5}$In$_{5}$, the bulk SC coexists {\em microscopically} 
with small-moment magnetism ($\leq$ 0.1$\mu$$_{B}$) \cite{LosAmos}. 
In all other heavy fermions, there are strong AF correlations present in 
the SC state \cite{Gabi1,Maple,LosAmos}. Another class of materials in 
which SC and AF coexist are cuprates such as YBCO and LSCO compounds 
\cite{Maple}. Coexistence of SC and FM order is found in the heavy fermion 
UGe$_{2}$ \cite{ferro}, and in Ru-based materials, for example, in 
RuSr$_{2}$GdCu$_{2}$O$_{8}$ \cite{Ru}.

In SC heavy-fermion systems, spin fluctuation (electron-electron 
interactions) are believed to mediate the electron pairing that leads to 
SC \cite{Varma}. For the heavy fermions CeIn$_{3}$, CePd$_{2}$Si$_{2}$ 
\cite{Martur}, UPd$_{2}$Al$_{3}$ \cite{Hunt} and UGe$_{2}$ \cite{ferro}, 
there is indirect evidence for spin-fluctuation mechanism of SC. This 
intimate relationship between the SC and magnetism also appears to be 
central to the SC cuprates \cite{Maple}, which inherited the magnetic 
properties from their parent compounds, AF Mott insulators. Many 
theoretical studies suggest that the SC in cuprates is mediated via the 
exchange of AF spin fluctuations \cite{Pine}. 

In cuprates, there are two energy scales \cite{Deu}: the pairing energy scale, 
$\Delta$$_{p}$ (see $\Delta$$_{sc}$ in Fig.7), and the phase-coherence 
scale, $\Delta$$_{c}$, observed experimentally \cite{Deu,AMour3}. The two 
energy scales have different dependences on hole concentration, $p$, in 
CuO$_{2}$ planes: $\Delta$$_{p}$ increases linearly with decrease in hole 
concentration, whereas $\Delta$$_{c}$ has approximately the parabolic 
dependence on $p$ and scales with $T_{c}$ as 
2$\Delta$$_{c}$ $\simeq$ 5.4$k_{B}T_{c}$ \cite{Deu}.
\begin{figure}[t]
\leftskip-10pt
\epsfxsize=0.9\columnwidth
\centerline{\epsffile{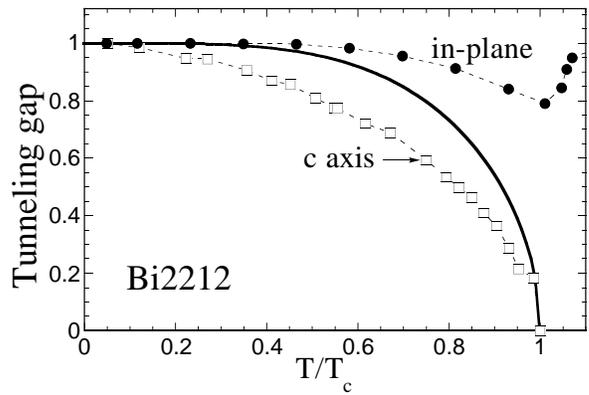}}
\vspace{2mm}
\caption{Temperature dependences of in-plane (dots) 
\protect\cite{Ekino1,Matsuda,AMour1,Oda} and $c$\,-axis 
tunneling QP peaks (squares) \protect\cite{Yurgens} in slightly overdoped 
Bi2212 single crystals, $\Delta$$(T)$/$\Delta$$(T_{min}$). The BCS 
temperature dependence is shown by the solid line. The dashed lines are 
guides to the eye.}
\label{fig12}
\end{figure}

\subsection{In-plane and $c$\,-axis tunneling in Bi2212}

Figure 12 shows the temperature dependence of in-plane tunneling QP 
peaks, measured in slightly overdoped Bi2212 single crystals 
\cite{Ekino1,Matsuda,AMour1,Oda}. 
Tunneling measurements performed on slightly underdoped Bi2212 single 
crystals show similar temperature dependence \cite{Miyakawa}. So, the 
temperature dependence of in-plane QP peaks shown in Fig.12 can be 
considered as typical. In Fig.12, we also present the temperature 
dependence of $c$\,-axis QP peaks measured in slightly overdoped 
Bi2212 mesas \cite{Yurgens}. Measurements performed on micron-size 
mesas present {\em intrinsic} properties of the material. In Fig.12, 
surprisingly, the temperature dependences along the $ab$ planes and along 
the $c$ axis are different. It is a clear hallmark of the coexistence of two 
different SC mechanisms in Bi2212: in-plane and along the $c$ axis.

\subsection{LSCO} 

In LSCO, there is evidence that the SC intimately relates to the 
establishment of AF order along the $c$ axis. Recent $\mu$SR 
measurements performed on non-SC Eu-doped LSCO having different hole 
concentrations show that the SC phase of pure LSCO is replaced in Eu-doped 
LSCO by the second AF phase (see Fig.4 in Ref.\cite{LSCO}). Thus, the data 
show that it is possible to switch the entire hole concentration dependent 
phase diagram from SC to AF. It is a clear hallmark that the SC in LSCO 
intimately relates to the formation of AF order. We return to this 
important result later.

We turn now to the analysis of resistivity data measured in non-SC 
2D layered compounds with AF or FM correlations. The data clearly 
show that, in all these layered compounds, the out-of-plane resistivity, 
$\rho$$_{c}$, has drastic changes either at N\'{e}el temperature, $T_{N}$, 
or Curie temperature, $T_{C}$, whereas the in-plane resistivity, 
$\rho$$_{ab}$, passes through $T_{N}$ or $T_{C}$ smoothly. 

We start with YBCO. In AF undoped YBCO (x = 0.35; 0.33; and 0.32) having 
$T_{N}$ $\simeq$ 80 K, 160 K, and 210 K, respectively, $\rho$$_{c}$ 
shows a sharp increase, by about 2 orders of magnitude, upon cooling through 
$T_{N}$ \cite{Lavrov1}. At the same time, $\rho$$_{ab}$ changes smoothly at 
$T_{N}$. The same effect has been observed in LuBCO (x = 0.34) 
\cite{Lavrov2}. So, the N\'{e}el ordering in undoped YBCO has 
remarkably different impact on the electron transport within CuO$_{2}$ 
planes and between them. The authors conclude \cite{Lavrov1}: 
``The N\'{e}el temperature actually corresponds to the establishment of AF 
order along $c$ axis.''

We now discuss the FM layered compounds. The structure of the FM compound 
Bi$_{2}$Sr$_{3}$Co$_{2}$O$_{9}$ (BSCoO) is similar to the structure of 
Bi2212, where CoO$_{2}$ planes are analogous with CuO$_{2}$ planes in 
Bi2212 \cite{Co}. BSCoO becomes FM at $T_{C}$ = 3.2 K. The resistivity 
data show that, at $T_{C}$, there is a cusp in $\rho$$_{c}$ but 
$\rho$$_{ab}$ changes smoothly. The authors conclude that the long-range 
magnetic order in BSCoO develops at $T_{C}$ along the $c$ axis \cite{Co}. 
In layered manganite La$_{1.4}$Sr$_{1.6}$Mn$_{2}$O$_{7}$ (LSMO) which 
is composed of the MnO$_{2}$ bilayers becomes FM at $T_{C}$ = 90 K 
\cite{MnO}. The resistivity data show that, at $T_{C}$, there are drastic 
changes in $\rho$$_{c}$ (a few orders of magnitude), but very small 
changes in $\rho$$_{ab}$. They conclude that, in LSMO, the long-range 
magnetic order develops at $T_{C}$ = 90 K along the $c$ axis \cite{MnO}.

Neutron-scattering measurements performed on the heavy fermion 
URu$_{2}$Si$_{2}$ ($T_{c}$ = 1.2 K) show that the AF order develops at 
$T_{N}$ = 17.5 K along the $c$ axis \cite{heavferm}. So, it seems that, in 
all layered compounds, the long-range AF or FM order develops at $T_{N}$ 
or $T_{C}$ along the $c$ axis (the in-plane magnetic correlations exist 
above $T_{N}$ and $T_{C}$ \cite{MnO}).

We return now to the analysis of the phase diagram of non-SC Eu-doped 
LSCO, where the SC phase of pure LSCO is replaced by the second AF 
phase \cite{LSCO}. The conclusion made in the previous paragraph 
signifies that either the main AF phase of Eu-doped LSCO or the second 
AF phase develops along the $c$ axis. Thus, the SC phase of pure LSCO is 
replaced in Eu-doped LSCO by the AF phase which develops along the $c$ 
axis. Consequently, the SC in LSCO intimately relates to the 
establishment of the long-range AF order along the $c$ axis.

In heavy fermion CeIrIn$_{5}$, $\mu$SR measurements discovered the 
onset of a small magnetic field ($\sim$ 0.4 Gauss) which sets exactly at 
$T_{c}$ \cite{LosAmos}. In YBCO (x = 0.6), recent INS measurements 
identified small magnetic moments directed along the $c$ axis, which 
increase in strength at and below $T_{c}$ \cite{Mook}.

\subsection{YBCO, Bi2212 and LSCO} 

Here we compare coherence SC and magnetic characteristics of YBCO, 
Bi2212 and LSCO. The comparison shows that the magnetic and coherence 
SC characteristics have similar temperature dependencies, and, at 
different dopings, their magnitudes are proportional to each other (and 
proportional to $T_{c}$). Thus, the coherence SC and magnetic properties 
of cuprates intimately relate to each other.

First, we describe the magnetic properties of cuprates. The low energy 
magnetic excitations in LSCO cuprate have been extensively studied, and 
the observed spin fluctuations are characterized by wave vector which 
is incommensurate with the lattice \cite{Lee}. These modulated spin 
fluctuations in LSCO persist in both normal and SC states. The spin 
dynamics in YBCO and Bi2212 studied by INS exhibit below $T_{c}$ a 
sharp commensurate resonance peak which appears at well defined 
energy $E_{r}$ \cite{i1,i2,i3,i4,i5,i6,i7,i8,i9,i10,i11}. 
Incommensurability in YBCO has been also reported \cite{i8}, and 
it is consistent with that in LSCO of the same hole doping, but, in YBCO, 
it occurs in the SC state. Now it is clear that the incommensurability 
and the commensurate resonance are inseparable parts of the general 
features of the spin dynamics in YBCO at all doping levels \cite{Mook2}.
 Thus, there is a clear evidence of coexistence of AF order and SC below 
 $T_{c}$, at least, in LSCO and YBCO.  

We now compare coherence SC and  magnetic characteristics of the 
cuprates. Figure 13(a) shows the temperature dependences of the 
superfluid density in near optimally doped single crystals of Bi2212 
\cite{rho1} and YBCO \cite{rho2}, and in an overdoped LSCO (x = 0.2) 
single crystal \cite{rho3}, measured by microwave, $\mu$SR and 
ac-susceptibility techniques, respectively. The superfluid density is 
proportional to 1/$\lambda$$^{2}(T)$, where $\lambda$$(T)$ is the 
magnetic penetration depth. Figure 13(b) shows the temperature 
dependences of Andreev-reflection gap measured in an overdoped Bi2212 
thin film \cite{andreev1}, and in overdoped single crystals of Bi2201 
\cite{andreev2} and LSCO (x = 0.2) \cite{andreev3}. It is important to 
emphasize that Andreev reflections are exclusively sensitive to 
coherence properties of condensate. In Figs 13(a) and 13(b), one can 
see that there is good agreement among temperature dependences of 
coherence SC characteristics of different cuprates. 

Figure 13(c) shows the temperature dependences of the peak intensity of 
the incommensurate elastic scattering in LSCO (x = 0) \cite{Lee} and the 
intensity of the commensurate resonance peak measured by INS in near 
optimally doped Bi2212 \cite{i1} and YBCO \cite{i6}. 

In Fig.13, all temperature dependences of coherence SC and magnetic 
characteristics below $T_{c}$ exhibit a striking similarity. Since 
all temperature dependences shown in Fig.13 are similar to the 
temperature dependence of $c$\,-axis QP peaks in Bi2212, shown in 
Fig.12, and different from the temperature dependence of in-plane QP 
peaks, also shown in Fig.12, it is reasonable to assume that the 
coherence SC and magnetic properties of the cuprates intimately relate 
to each other along the $c$ axis. Moreover, in YBCO (x = 0.6), recent INS 
measurements found small magnetic moments directed along the $c$ axis, 
which increase in strength at and below $T_{c}$ \cite{Mook}. At the same 
time, as one can see in Figs 12 and 13, the in-plane mechanism of the SC 
in the cuprates has no or little relations to the magnetic interactions, at
least, along the $c$ axis. 
\begin{figure}[t]
\leftskip-10pt
\epsfxsize=0.8\columnwidth
\centerline{\epsffile{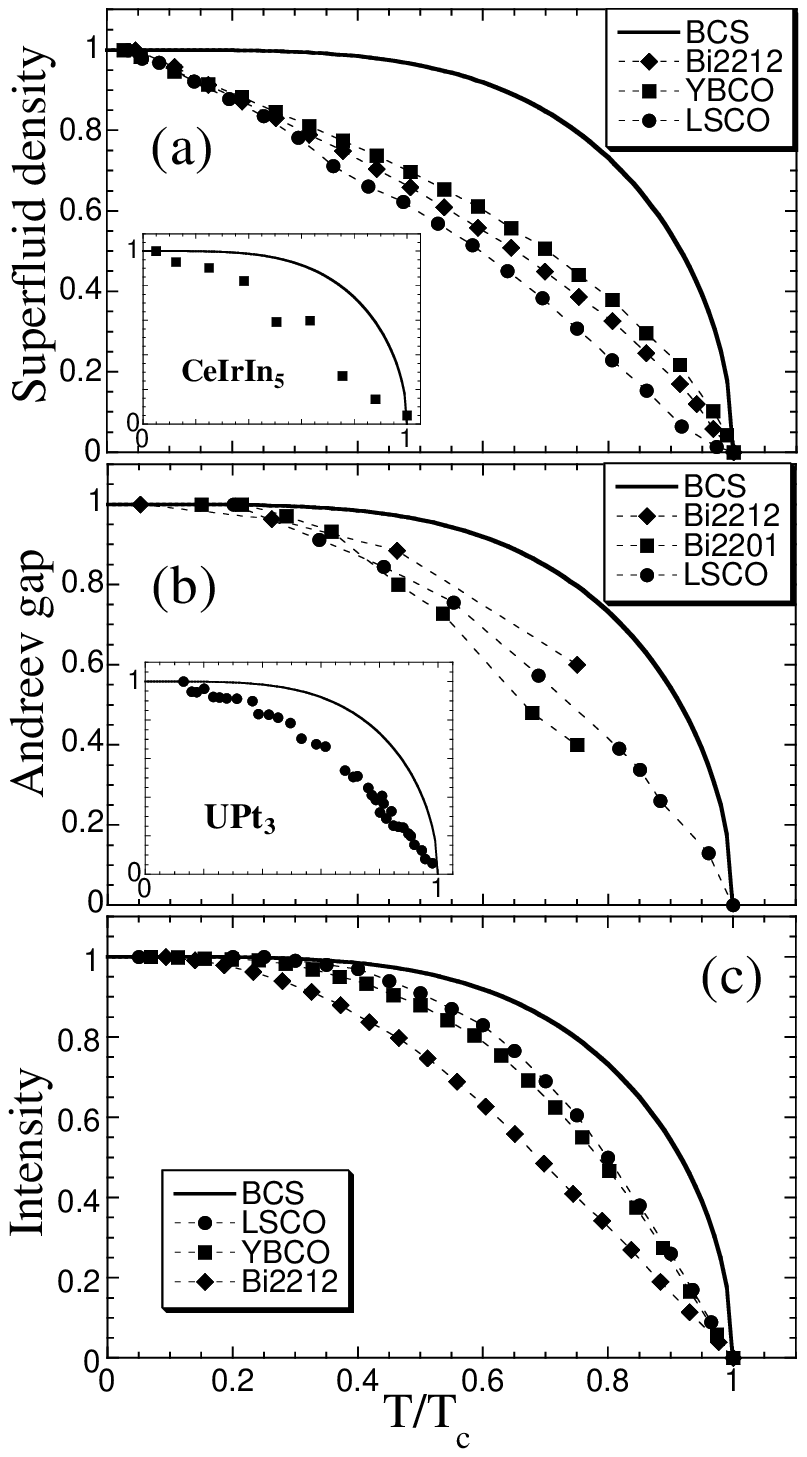}}
\end{figure}

Figure 14 shows the energy position of the magnetic resonance peak, 
$E_{r}$, in Bi2212 \cite{i1,i2} and in YBCO 
\cite{i3,i4,i5,i6,i7,i8,i9,i10,i11} as a function of doping. 
The parabolic curve corresponds to the coherence energy scale, 
$\Delta$$_{c}$, which is proportional to $T_{c}$ as 
2$\Delta$$_{c}$ $\simeq$ 5.4$k_{B}T_{c}$ \cite{Deu}. At 
different doping levels, Andreev-reflection data coincide with 
$\Delta$$_{c}$ \cite{Deu,AMour3,andreev3}. 
In Fig.14, we present also the $\Delta_{sc}$ data from Fig.7, which
correspond to the pairing energy scale \cite{Deu}. In Fig.14, one 
can see that, at different dopings, $E_{r}$ is proportional to 
$\Delta$$_{c}$ as $E_{r}$ $\simeq$ 2$\Delta$$_{c}$. 
This correlation suggests that spin excitations are responsible for 
establishing the phase coherence in YBCO and Bi2212 since the relation 
$E_{r}$ $\simeq$ 2$\Delta$$_{c}$ is in good agreement with the 
theories in which the SC is mediated by spin fluctuations \cite{Pine}.
Recently, it was shown that applied magnetic fields suppress the 
magnetic resonance peak in YBCO, indicating that the resonance peak
indeed measures the long-range phase coherence \cite{Dai}. The 
strength of coupling between spin excitations and charge carriers is 
sufficient to account for the high $T_{c}$ value in cuprates \cite{Basov}.

What is interesting is that all temperature dependences shown in Fig.13
are similar to the temperature dependence of the superfluid density in 
heavy fermion CeIrIn$_{5}$, measured by $\mu$SR\cite{LosAmos}, and to 
the temperature dependence of the Andreev-reflection gap in heavy 
fermion UPt$_{3}$ \cite{DeWilde}, which are shown in the insets of 
Figs 13(a) and 13(b), respectively. Spin fluctuations are believed to 
mediate the electron pairing in CeIrIn$_{5}$ \cite{LosAmos} and UPt$_{3}$ 
\cite{Varma} that leads to SC. The magnetic resonance peak has not yet 
been detected in CeIrIn$_{5}$ or UPt$_{3}$, however, the magnetic 
resonance peak has been observed by INS in another heavy fermion 
UPd$_{2}$Al$_{3}$ \cite{Metoki} where spin fluctuations mediate the SC 
which coexists with the long-range AF order \cite{Hunt}. The latter facts 
also point to the presence of the spin-fluctuation coupling mechanism in 
cuprates. 

The behavior of all temperature dependences shown in Fig.13 can be 
easily understood in terms of the spin-fluctuation SC mechanism 
(electron-electron interactions): crudely speaking, they exhibit the 
squared BCS temperature dependence.
\begin{figure}[t]
\caption{(a) Temperature dependence of the superfluid density in near 
optimally doped single crystals of Bi2212 ($T_{c}$ = 93 K) 
\protect\cite{rho1} and YBCO ($T_{c}$ = 93 K) \protect\cite{rho2}, and 
in an overdoped LSCO (x = 0.2) single crystal ($T_{c}$ = 36 K) 
\protect\cite{rho3}. The superfluid density is 
proportional to 1/$\lambda$$^{2}(T)$, where $\lambda$$(T)$ is the 
magnetic penetration depth. Inset: temperature dependence of the 
superfluid density (squires) in heavy fermion CeIrIn$_{5}$ 
($T_{c}$ = 0.4 K) \protect\cite{LosAmos} (axis parameters as main plot). 
(b) Temperature dependence of Andreev-reflection gap, 
$\Delta$$(T)$/$\Delta$($T_{min}$), in an overdoped Bi2212 thin film 
($T_{c}$ = 80 K) \protect\cite{andreev1}, and in overdoped single crystals 
of Bi2201 ($T_{c}$ = 29 K) \protect\cite{andreev2} and LSCO (x = 0.2) 
($T_{c}$ = 28 K) \protect\cite{andreev3}. Inset: temperature dependence of 
the Andreev gap (dots) in heavy fermion UPt$_{3}$ ($T_{c}$ $\sim$ 440 mK) 
\protect\cite{DeWilde} (axis 
parameters as main plot). (c) Temperature dependence of the peak 
intensity of the incommensurate elastic scattering in LSCO (x = 0) 
($T_{c}$ = 42 K) \protect\cite{Lee} and the intensity of the magnetic 
resonance peak measured by INS in near optimally doped Bi2212 
($T_{c}$ = 91 K) \protect\cite{i1} and YBCO ($T_{c}$ = 92.5 K) 
\protect\cite{i6}. The neutron-scattering data are average, the real data 
have the vertical error of the order of $\pm$10\% \protect\cite{Lee,i1,i6}. 
The BCS temperature dependence is 
shown by the thick solid line. The dashed lines are guides to the eye.}
\label{fig13}
\end{figure}

\subsection{Discussion} 

In spite of the unmistakable similarities among the magnetic and SC 
properties of YBCO, Bi2212 and LSCO (and some heavy fermions for 
which there is an indirect evidence for spin-fluctuation mechanism of 
SC), clearly, there is a difference between magnetic properties of LSCO 
and YBCO. If, in YBCO, $T_{c}$ = $T_{com}$ = $T_{inc}$, where $T_{com}$ 
($T_{inc}$) is the onset temperature of the (in-) commensurate peak(s), 
in LSCO the situation is different. First, the commensurate peak has 
not been detected. Second, in LSCO, mainly, $T_{c} < T_{inc}$ 
(Figure 13(c) shows the case when $T_{c}$ = $T_{inc}$). So, the magnetic 
and SC properties of LSCO are similar to those of most heavy fermions: 
the commensurate peak has not been detected, and $T_{c} < T_{N}$. 
For example, in UPd$_{2}$Al$_{3}$, $T_{c}$ $\approx$ 2 K and 
$T_{N}$ = 14.5 K \cite{Hunt,Metoki}; in URu$_{2}$Si$_{2}$, 
$T_{c}$ $\approx$ 1.2 K and $T_{N}$ = 17.5 K \cite{heavferm}, and, in 
FM heavy fermion UGe$_{2}$, $T_{c} < T_{C}$ \cite{ferro}. In Bi2212, the 
situation looks more like that of YBCO, even if, the incommensurate 
peaks have not yet been detected. 

In fact, the differences between magnetic and SC properties of YBCO and 
LSCO (and among some heavy fermions) can be understood in terms of the 
following chain of events: the formation of pairs at $T_{pair}$ - the
SC order parameter couples to the magnetic in-plane order parameter - 
the appearance of AF interactions along the $c$ axis at $T_{m}$ - the 
appearance of the long-range SC phase coherence. If there is no pairs, the 
SC is absent, even if, the AF order is established (as in the Eu-doped 
LSCO \cite{LSCO}). If $T_{pair} < T_{m}$, then, this reminds the situation in 
LSCO where $T_{c} < T_{inc}$. If $T_{pair} > T_{m}$, then, this reminds the 
case of YBCO where $T_{c}$ = $T_{com}$ = $T_{inc}$, and, consequently, 
$T_{c}$ = $T_{m}$. Unfortunately, this simple picture cannot explain the 
presence of the commensurate peak in INS spectra but, at least, shows 
that the differences between the magnetic and SC properties of LSCO and 
YBCO can be understood in the framework of one picture with different 
initial parameters.

\subsection{Summary}
 
Analysis of the data obtained by different techniques in 
YBCO, Bi2212 and LSCO shows that the long-range phase coherence 
intimately relates to AF interactions along the $c$ axis. Apparently, 
in cuprates, the magnetic and SC order parameters are coupled to 
each other, and the phase-coherence scale, $\Delta$$_{c}$, has the 
magnetic origin (see Fig.14).

There are common features in the SC state of the heavy fermions 
CeIrIn$_{5}$, UPt$_{3}$ and UPd$_{2}$Al$_{3}$, on the one hand, and the 
cuprates, on the other hand. It is possible that, in all heavy-fermion and 
organic SCs, the long-range phase coherence is established due to spin 
fluctuations. The pairing mechanism may be different (as in Bi2212).

\section{Tunneling assisted by spin excitations in Bi2212}

In Fig.14, one can see that at $p$ = 0.2, 
$E_{r} \simeq 2\Delta_{sc}$. Therefore, the tunneling in a 
{\em SIS}\, junction of Bi2212 having $p$ $\simeq$ 0.2 can be assisted 
by spin excitations \cite{Tsui}. Bearing this fact in mind, we 
interpret the peak at 35 mV in the difference shown in Fig.9(a) as the 
contribution from electron tunneling assisted by spin excitations.
Indeed, the peak position in Fig.9(a), 35 mV, and the magnitude of SG 
in Fig.5(b), 2$\Delta$$_{sc}$ = 34 meV, have similar values.
\begin{figure}[t]
\leftskip-10pt
\epsfxsize=0.9\columnwidth
\centerline{\epsffile{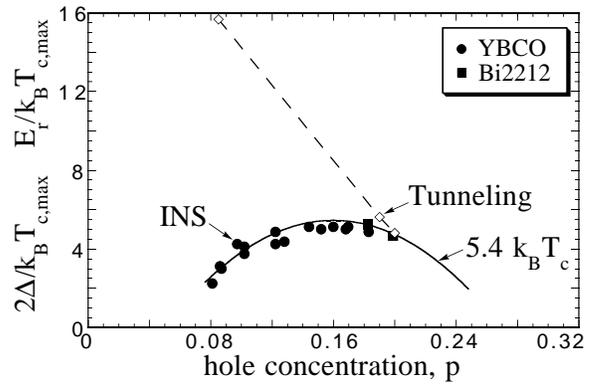}}
\vspace{2mm}
\caption{Phase diagram: the energy position of the magnetic 
resonance peak, $E_{r}$, in Bi2212 (squares) \protect\cite{i1,i2} and 
in YBCO (dots) \protect\cite{i3,i4,i5,i6,i7,i8,i9,i10,i11} and 
$\Delta_{sc}$ (diamonds) from Fig.7. The parabolic solid line corresponds 
to the coherence energy scale, $\Delta$$_{c}$ \protect\cite{Deu}, which 
is proportional to $T_{c}$ as 2$\Delta$$_{c}$ $\simeq$ 5.4$k_{B}T_{c}$. 
When not available, $p$ has been calculated from the relation 
$T_{c}$/$T_{c,max}$ = 1 - 82.6($p$ - 0.16)$^{2}$ \protect\cite{Tallon}, 
and we use $T_{c,max}$ = 93 K and $T_{c,max}$ = 95 K in the case of YBCO 
and Bi2212, respectively. The dash line is a guide to the eye.}
\label{fig14}
\end{figure}

If the interpretation is correct, then the tunneling data in Fig.5(b)
present additional evidence that spin fluctuations mediate
the long-range phase coherence in Bi2212. First, this shows 
that charge carriers are strongly coupled to spin excitations. Second,
by comparing the values of the Josephson $I_{c}R_{n}$ product of
the spectra shown in Figs 4(b) and 5(b) (7.5 mV and 24.5 mV, 
respectively), the high value of the Josephson product of the spectra in 
Fig.5(b) can only occur due to the contribution from the tunneling 
assisted by spin excitations because this contribution is 
the {\em only} difference between the spectra shown in Figs 4(b) and 
5(b) [see Figs 8(c) and 9(a)]. Consequently, this signifies that the 
spin fluctuations mediate the phase coherence in Bi2212.

\section{The two energy scales}

The two energy scales shown in Fig.14, both relate to the SC: the linear 
$\Delta$$_{sc}$ (or $\Delta$$_{p}$) scale is the in-plane energy scale, 
whereas the scale proportional to 
$T_{c}$ is the $c$\,-axis energy scale, and they have different origins.
It is important to emphasize that the $\Delta$$_{sc}$ energy scale
is observed by tunneling spectroscopy and ARPES, whereas the $\Delta$$_{c}$ 
energy scale is measured in Andreev reflections \cite{Deu}. However, in some
cuprates, the $\Delta$$_{c}$ scale is also observed in tunneling measurements 
along the $c$ axis \cite{AMour3,AMour8,Wei}. For example, in near optimally 
doped YBCO single crystals, the maximum magnitudes of the tunneling gap
into CuO$_{2}$ planes and along the $c$ axis are 28 meV and 19 meV,
respectively \cite{Wei}. The same values measured in electron-doped 
Nd$_{1.85}$Ce$_{0.15}$CuO$_{4-\delta}$ (NCCO) underdoped in oxygen are 
13 meV and 3.5 meV, respectively, and the magnitude of the Andreev gap is
3.5 meV \cite{AMour3}. So, in YBCO and NCCO, tunneling measurements along 
the $c$ axis show the $\Delta$$_{c}$ scale. However, it is not the case in 
Bi2212, where tunneling measurements along the $c$\,-axis show the 
$\Delta$$_{sc}$ scale \cite{Yurgens,Rener}.
 
The symmetries of the two energy scales are discussed elsewhere 
\cite{AMour9}: the pairing energy scale, $\Delta$$_{sc}$ has most likely 
an anisotropic s-wave symmetry like the PG (CDW gap) which defines the 
magnitude of $\Delta$$_{sc}$, whereas the coherence energy scale, 
$\Delta$$_{c}$, has the d-wave symmetry. Since spin fluctuations 
mediate the phase coherence, they are responsible for the d-wave 
symmetry of $\Delta$$_{c}$ \cite{Pine}. All phase-sensitive techniques, 
obviously, probe the symmetry of $\Delta$$_{c}$, and not of 
$\Delta$$_{sc}$, thus they detect the d-wave symmetry. At the same time, 
tunneling measurements show a s-wave symmetry of the condensate 
\cite{Klem2}, {\em i.e.} the symmetry of $\Delta$$_{sc}$.

\section{The anisotropy of the tunneling PG}

Let us concentrate for a while on the tunneling PG. The analysis of 
the data (see above) suggests that the tunneling PG is predominantly 
a charge gap on charge stripes in CuO$_{2}$ planes. This charge gap is most 
likely a CDW gap. It is not difficult to show the shape of the CDW gap. As 
mentioned above, the shape of the PG shown in the inset of Fig.4(b) indicates 
that the PG has most likely an anisotropic s-wave symmetry. Second, from 
the data presented here and in Ref.\cite{Miyakawa}, the maximum 
magnitudes of in-plane PG and SG in Bi2212 correlate with each other 
as $\Delta_{ps,in} \simeq 2\Delta_{sc,in}$. The intrinsic $c$\,-axis 
tunneling data in Bi2212 mesas show that, at low temperature, 
$\Delta_{ps,c} \simeq \Delta_{sc,c}$ (see Fig.1 in Ref.\cite{Yurgens2}). 
Since $\Delta_{sc,c} \simeq \Delta_{sc,in}$ (compare our data 
with the data in the overdoped sample in Ref.\cite{Yurgens}), 
it is reasonable to assume that the value of $\Delta_{ps,c}$ 
corresponds to the minimum of the in-plane CDW gap because 
the tunneling along the $c$\,-axis is {\em preferable} from the 
minimum of in-plane CDW gap. Then, one can conclude that the 
in-plane CDW gap in Bi2212 is anisotropic with the anisotropy 
ratio of $\Delta_{ps,max}$/$\Delta_{ps,min} \simeq$ 2. 

From quasi-1D structure of the stripes, the CDW gap has either two- 
or four-fold symmetry. The maximum magnitude of the SG is most 
likely defined by the minimum of the charge gap on the stripes. 

\section{Changes in Bi2212 with decrease of temperature}

By synthesizing our findings we briefly describe here the changes 
occurring in Bi2212 with decrease in temperature.

Analysis of the data shows that (i) the tunneling characteristics 
corresponding to the QP peaks in Bi2212 are in excellent agreement 
with theoretical predictions made for a quasi-1D topological-excitation 
liquid, and (ii) the phase coherence along the $c$\,-axis in cuprates is 
established at $T_{c}$ due to spin fluctuations in local AF domains of 
CuO$_{2}$ planes. By taking into account (i), we assume that the quasi-1D 
topological excitations reside on quasi-1D charge stripes. The assumption 
is reasonable because AF domains which separate the quasi-1D charge 
stripes are truly 2D \cite{Tranquada}. Consequently, the quasi-1D 
excitations cannot reside into the AF domains.

Depending on the compound, the charge stripes are formed in cuprates
at $T_{charge} \sim$ 70--300 K (or higher) (see references in 
Ref.\cite{AMour2}). The excitations along stripes appear with 
appearance of the stripes, thus, at $T_{charge}$. The stripe 
excitations couple with each other at $T_{onset} < T_{charge}$, which 
can be considered as an onset of SC in Bi2212. For example, in slightly 
overdoped Bi2212, two independent tunneling studies 
\cite{Ekino1,AMour1} show that the onset of SC occurs at 110--116 K. 
Finally, the long-range phase coherence is established at $T_{c}$ 
along the $c$\,-axis ({\it i.e.} between CuO$_{2}$ planes) due to spin 
fluctuations. Thus, there are 3 characteristic temperatures: 
$T_{c}$ $\leq$ $T_{onset} < T_{charge}$. In addition, between $T_{onset}$ 
and $T_{charge}$, there is a characteristic temperature 
\cite{Ekino1,AMour1} which was attributed to the spin ordering in AF 
domains \cite{AMour1}.

The critical temperature, $T_{c}$, can be measured directly, and 
$T_{charge}$ and $T_{onset}$ are proportional to $\Delta$$_{ps}$ and 
$\Delta$$_{sc}$ (see Fig.7), respectively. It is interesting to note that, 
since $\Delta_{ps}$ $\simeq$ 2$\Delta_{sc}$, $T_{onset}$ is always about 
twice as smal as $T_{charge}$. It is possible to estimate the relation 
between $T_{onset}$ and $\Delta$$_{sc}$. By using the data measured in 
slightly overdoped Bi2212 and presented in Refs \cite{Ekino1,AMour1}, we 
have $T_{onset}$ $\approx$ 0.4$\Delta$$_{sc}$/$k_{B}$. 

Earlier, we proposed a magnetic coupling between stripes (MCS) model
assuming that the Cooper pairs are formed along charge stripes,
and the phase coherence is established due to magnetic coupling 
between SC stripes \cite{AMour6}. From the present work, the general 
idea of the MCS model is correct, however, the more precise name of 
the model of SC in cuprates should be the magnetic coupling between 
striped CuO$_{2}$ planes.

\section{Cooper pairs}

As discussed in Section XI, the topological solitons condense at 
$T_{onset}$ by lowering their energy. Formally, the topological excitations 
shown in Fig.11 are ''pure'' charge excitations. The question is at what 
temperature is the spin sector condensed, at $T_{onset}$ or $T_{c}$? If 
Davydov's {\em bisoliton} scenario \cite{Davydov1,Davydov2} is realized 
than the spins condense together with the charges at $T_{onset}$ by 
creating singlet states. If the condensate consists of kink-kink bound 
states, then spins may condense at $T_{c}$. Since ''pure'' charge 
excitations have boson-like properties, generally speaking they do not need 
to create a bound state. So, it is {\em most likely} that the Cooper 
pairs in Bi2212 are Davydov's bisolitons, and the phase coherence among the
bisolitons is established at $T_{c}$ due to spin fluctuations (excitations). 

It is important to emphasize that, in such a scenario, formally, the Cooper 
pairs exist above $T_{c}$, between $T_{c}$ and $T_{onset}$. However, 
{\it de fait}, they are the {\em local} Cooper pairs: In order to jump 
from one stripe to another (above $T_{c}$), they have similar 
difficulties as a single electron.

One would be interested to know what is the glue between two electrons
in the soliton scenario for the SC in Bi2212. It is phonons, and the 
electron-phonon interactions are strong and nonlinear 
\cite{Davydov1,Davydov2}.

\section{Data obtained in cuprates}

The quasi-1D topological-excitation-liquid scenario is attractive, 
because by using it, it is possible to explain experimental data 
obtained in cuprates. Let's {\em briefly} discuss {\em some} data. 

The discrepancy for the presence of the PG in the overdoped 
region in transport and tunneling measurements can be easily 
understood: they measure two different PGs. 
In tunneling measurements, it is predominantly the charge gap, and,
in transport measurements, it is the resistance to the 
propagation of stripe excitations along stripes which 
are located in AF environment. For example, if a molecular chain is 
embedded into a medium, a frictional force acts on the soliton
\cite{Davydov1,Davydov2}.

In frameworks of the quasi-1D topological-excitation-liquid scenario 
in Bi2212, magnetic and non-magnetic impurities will affect the SC 
similarly, which is the case in cuprates \cite{Dow}.
The oxygen isotope effect \cite{Muller2} can be naturally understood 
since the stripe dynamics above $T_{c}$ depends on the underlying 
lattice \cite{Bianconi}. At the same time, a very weak ''BCS isotope
effect'' in cuprates is well understood in terms of the soliton SC 
\cite{Davydov1,Davydov2}.

Mysterious vortex-like excitations found in the Nernst effect 
\cite{Ong} correspond to the stripe excitations. The authors are correct 
by pointing out that the observed excitations are vortex-like because 
vortices themselves are solitons \cite{french}.
Such scenario in Bi2212 can explain why the phenomenon of SC in hole- 
and electron-doped cuprates can be understood within a common 
scheme \cite{AMour3}. The occurrence of sudden 
CDW ordering at low temperature in undoped cuprates \cite{AMour2} 
and the insulating behavior in a strong magnetic field \cite{Ando}
correspond to the disappearance of the stripe excitations.

By using the quasi-1D topological-excitation-liquid 
scenario in Bi2212 and the fact that the phase coherence along 
the $c$\,-axis is established due to spin fluctuations,
it is possible to explain why $T_{c}$ and 
$\Delta_{sc}$ do not correlate with each other in cuprates:
the magnitude of $\Delta_{sc}$ is defined by the 
magnitude of the charge gap on stripes, at the same time, the value of
the {\em bulk} $T_{c}$ value depends on spin fluctuations into 
local AF domains. 

It is possible that zero-bias conductance peak observed in 
tunneling measurements \cite{AMour3,Wei} corresponds not only to
Andreev surface bound states \cite{Hu1,Tanaka} but also to the
topological solitons [see Fig.2(a)]. 

The SC condensation energy has the maximum in the overdoped region
near $p$ = 0.19 \cite{Tallon,Shen}. Why? The SC in Bi2212 is literally 
spread into the two channels which have different energy scales: 
in-plane and along the $c$\,-axis. From Fig.14, the two energy
scales have the same values approximately at $p$ = 0.2. Thus, the two
energy scales emerge into one at $p$ = 0.19--0.2.

Recently, a new energy scale is found in ARPES spectra of hole-doped 
cuprates: an abrupt change (a kink) has been observed in the electronic QP 
dispersion \cite{Pasha}. The kink is associated either with phonons or with
the magnetic resonance peak. In fact, the clue to the origin of the kink
in ARPES spectra can be found in Ref.\cite{Valla}. Measurements in 2D 
layered 2H-TaSe$_{2}$ CDW compound found a similar energy scale in 
ARPES spectra \cite{Valla}. The authors conclude \cite{Valla}: 
''A reduction in the scattering rates below this energy indicates the 
collapse of a major scattering channel with the formation of the CDW
state accompanying the appearance of a bosonic mode in the excitation
spectrum of the system.'' In the CDW compound, it is clear that this 
bosonic mode corresponds to the appearance of solitons in the CDW state. 
Consequently, the origin of the kink in ARPES spectra of hole-doped 
cuprates is the same, {\em i.e.} due to solitons. In fact, it is also 
possible that this kink corresponds to the {\em real} Fermi surface in 
cuprates and in 2H-TaSe$_{2}$ (see Fig.2 in Ref.\cite{Mele}).

Lastly, it is important to note that, in the framework of the quasi-1D 
topological-excitation-liquid scenario in Bi2212, one can explain the 
violation of the sum rule, observed in infrared measurements 
\cite{Basov2}: the stripe excitations do not obey the sum rule.

\section{Conclusions}

We would like to emphasize that we do not present here an
evidence that the tunneling pseudogap in Bi2212 is a CDW gap. 
This conclusion is based on the analysis of the data.
Here we found that the tunneling characteristics corresponding to 
quasiparticle peaks are in excellent agreement with theoretical 
predictions made for a quasi-1D topological-excitation liquid.
In fact, this information is sufficient in order to conclude that the 
tunneling pseudogap in Bi2212 is predominantly a charge gap. Since
solitonlike excitations, and not electrons (holes), play the role of 
quasiparticles in Bi2212, this implies that the stripes are 
{\em insulating}. In other words, there is a charge gap along stripes which
is most likely a CDW gap. Then, in order to pull out an electron (a hole) 
from a stripe, it is necessary, first of all, to overcome the charge 
gap on the stripe. Consequently, the tunneling pseudogap is 
predominantly the charge gap. However, it is possible that the pseudogap 
shown in Fig.3(b) consists of two (or a few) contributions: 
the predominant contribution from the charge gap and another contribution, 
for example, from AF correlations.

In summary, tunneling measurements have been carried out on 
underdoped, overdoped and Ni-substituted Bi2212 single crystals. 
Tunneling spectra below $T_{c}$ are the combination of 
incoherent part from the pseudogap and coherent 
quasiparticle peaks. There is a clear correlation between the 
maximum magnitude of the pseudogap and the distance between the 
quasiparticle peaks. The tunneling characteristics corresponding to the
quasiparticle peaks are in excellent agreement with theoretical 
predictions made for a quasi-1D topological-excitation liquid.
So, the quasiparticle peaks in Bi2212 appear from the 
condensation of quasi-1D topological excitations which reside 
most likely on charge stripes. Analysis of the data suggests that (i) 
the {\em tunneling} pseudogap in Bi2212 is predominantly a charge gap 
(CDW gap) on dynamical charge stripes, which has most likely an anisotropic 
s-wave symmetry, and (ii) the Cooper pairs in Bi2212 are Davydov's 
bisolitons. In addition, analysis of data 
measured by different techniques shows that the phase coherence is 
established at $T_{c}$ due to spin fluctuations in local antiferromagnetic 
domains of CuO$_{2}$ planes. It is possible that the magnon-like excitations 
which cause the appearance of the magnetic resonance peak in inelastic 
neutron scattering spectra are in resonance with the bound states of 
solitons \cite{Buryak}.

Many other data obtained in cuprates can be naturally understood 
in the framework of the quasi-1D topological-excitation-liquid scenario.
In the phase diagram of cuprates, the topological-excitation-liquid 
conductor known as a "strange metal" is located between an
antiferromagnetic Mott insulator at low hole concentration and a 
Fermi-liquid metal at high hole concentration. 

\acknowledgments
I would like thank D. Davydov, A. V. Buryak, K. Maki 
and M. V. Mostovoy  for discussions, 
N. Miyakawa for sending the data from Ref.\cite{Miyakawa2}, 
and many participants of the "Stripe 2000" Conference in Rome, with
whom I discussed the results of the present paper.

\end{document}